\documentclass[english,prd,twocolumn,footinbib]{revtex4}
\pdfoutput=1
\usepackage{graphicx, bm, color, babel}
\usepackage{hyperref}
\usepackage{amsmath}
\usepackage{color}

\def\be{\begin{equation}}
\def\ee{\end{equation}}
\def\bea{\begin{eqnarray}}
\def\eea{\end{eqnarray}}

\def\d {\mathrm{d}}
\newcommand{\HH}{\mathcal{H}}
\newcommand{\D} {\nabla}

\renewcommand{\>}{\rangle}

\newcommand{\p}{_{{\text{\tiny$\|$}}}}
\newcommand{\pp}{{{\text{\tiny$\|$}}}}
\renewcommand{\o}{{\hspace{-0.5mm}{\text{\tiny$\perp$}}}}

\newcommand{\De}{\Delta}
\newcommand{\de}{\delta}
\newcommand{\Om}{\Omega}
\newcommand{\omm}{\omega_m}
\newcommand{\omb}{\omega_b}
\newcommand{\La}{\Lambda}
\newcommand{\la}{\lambda}
\newcommand{\dd}{\partial}

\newcommand{\bn}{{\bm n}}

\begin{document}

\title{What is the distance to the CMB?}

\author{Chris Clarkson$^1$, Obinna Umeh$^{2}$,  Roy Maartens$^{2,3}$ and Ruth Durrer$^4$\\[3mm]
\emph{$^1$Astrophysics, Cosmology \& Gravity Centre, and, Department of Mathematics \& Applied Mathematics, University of Cape Town, Cape Town 7701, South Africa.\\
$^2$Physics Department, University of the Western Cape,
Cape Town 7535, South Africa \\
$^3$Institute of Cosmology \& Gravitation, University of Portsmouth, Portsmouth PO1 3FX, United Kingdom\\
$^4$D\'epartement de Physique Th\'eorique \& Center for Astroparticle Physics, Universit\'e de Gen\`eve, CH-1211 Gen\`eve 4, Switzerland.
}}

\begin{abstract}

The success of precision cosmology depends not only on accurate observations, but also on the theoretical model -- which must be understood to at least the same level of precision. Subtle relativistic effects can lead to biased measurements if they are neglected.  
One such effect gives a systematic shift in the distance-redshift relation away from its background value, due to the non-linear relativistic conservation of total photon flux. We also show directly how this shift
follows from a fully relativistic analysis of the geodesic deviation equation.  
We derive the expectation value of the shift using second-order perturbations about a concordance background, and show that 
the distance to last scattering is increased by 1\%. We argue that neglecting this shift could lead to a significant bias in the background cosmological parameters, because it alters the meaning of the background model. 
A naive adjustment of CMB parameter estimation if this shift is really a correction to the background would raise the $H_0$ value inferred from the CMB by 5\%, potentially removing the tension with local measurements of $H_0$. Other CMB parameters which depend on the distance would also be shifted by $\sim$1$\sigma$ when combined with local $H_0$ data. While our estimations rely on a simplistic analysis, they nevertheless illustrate that accurately defining the background model in terms of the expectation values of observables is critical when we aim to determine the model parameters at the sub-percent level. 

\end{abstract}

\maketitle

\section{Introduction}

Cosmology has entered a precision era. The premier cosmological dataset is the anisotropies and polarization of the cosmic microwave background (CMB). This is not only due to the highly accurate data, but also because of its simple  theoretical description, which allows accurate calculations. Present CMB codes like CAMB~\cite{Lewis:1999bs} and CLASS~\cite{Blas:2011rf} are typically 0.1\% accurate and, together with contemporary data, provide a determination of basic cosmological parameters to the percent level (and substantially better for the case of curvature)~\cite{Ade:2013zuv}. A puzzling result from current CMB measurements is that the Hubble parameter $H_0$ is significantly smaller than the value measured locally~\cite{Riess:2011yx,Marra:2013rba,Busti:2014dua}. 

Parameter estimation from the  CMB is extremely sensitive to the angular diameter distance $d_A(z_*)$, where  $z_*\simeq 1090$ is the redshift of the last scattering surface. More precisely it depends on the angular size of the sound horizon, $\theta = r_*/d_A(z_*)$, where $r_*$ is the sound horizon at last scattering. The
Planck collaboration~\cite{Ade:2013zuv} has reported $\theta =(1.04131\pm 0.00063)\times 10^{-2}$; hence it measures this scale with an accuracy of better than $10^{-3}$. The accuracy of $r_*$ is slightly worse, about $4.3\times10^{-3}$, which is also the accuracy of $d_A(z_*)$. These numbers indicate that a change of $\sim 1$  percent in $d_A(z_*)$ is critical for parameter estimation of the CMB at the present level of accuracy.

Most calculations of the CMB anisotropies are  performed within first-order perturbation theory and only CMB lensing requires a second-order analysis including first-order angular deflection of first-order temperature anisotropies and polarization. In this work we consider  the homogeneous change in the angular diameter distance due to the presence of structures in the Universe, up to second order in perturbation theory. 
We argue that it is critical to include this change at the present level of accuracy, as it induces a change in the mean distance to the last scattering surface which is larger than the current measurement error. 
Including this change could remove the the tension between the values of $H_0$ obtained by CMB observations and by local measurements. This change then affects other parameters which depend on $H_0$. For example,  $\omm=\Om_mh^2$ is well determined by the CMB, and so a change in  $h$ implies $\Omega_m$ moves significantly away from its naive value inferred without inclusion of this relativistic second-order correction. 
(Here $h$ is defined by $H_0= 100h\,$km\,s$^{-1}$Mpc$^{-1}$.) 
 
The observed angular diameter distance at observed redshift $z_s$ in direction $\bm n$ is 
\bea\label{bdjscksjdb}
d_A(z_s,\bm n)&=&{1\over 1+z_s}\left[\chi_s + \delta d_A(z_s,\bm n)+\frac{1}{2}\delta^2 d_A(z_s,\bm n)\right]\nonumber\\
&=&\bar{d}_A(z_s)[1+\Delta(z_s,\bm n)], \label{da}
\eea
which has a perturbation $\Delta(z_s,\bm n)$ about the 
 the background distance
 \begin{equation}
 \bar{d}_A(z_s) =\frac{\chi_s}{(1+{z}_s)}= \frac{1}{(1+{z}_s)}\int_{0}^{{z}_s} \frac{\d z}{(1+z)\HH(z)}
  \,.
\end{equation}
Here $\chi_s$ is the comoving distance (in the background geometry) to the source at observed redshift $z_s$ and $\HH$ is the conformal Hubble rate.  
The perturbation $\Delta(z_s,\bm n)$ arises from the fact that the Universe is not homogeneous and isotropic, but contains cosmic structures which induce fluctuations in the geometry. At linear order in perturbation theory, the lensing convergence $\kappa=-\Delta$  produces no change in the mean value $\langle{d}_A\rangle$ (although it does give a variance)~\cite{Bonvin:2005ps}.  At second order, however,  nonlinear effects give a relativistic correction to the distance-redshift relation that a typical observer would expect. 

This correction can be calculated from the ensemble average:
\be\label{ens}
d^\text{eff}_A(z_s) = \langle d_A(z_s,\bm n)\rangle=\bar{d}_A(z_s)[1+\langle\Delta\rangle\!(z_s)],
\ee
where we assume statistically isotropic Gaussian initial perturbations, so that there is no dependence on directions (all directions receive the same correction). If it is not correctly taken into account, the shift in the `background' distance-redshift relation by $\langle\Delta\rangle$ could result in a shift in the inferred cosmological parameters which appear in the distance-redshift relation. We begin by anticipating that this correction is imposed by conservation of flux, and we show rigorously in the appendices that this follows from purely geometrical arguments using the geodesic deviation equation for null geodesics. 

\section*{Flux conservation requires a shift in angular diameter distance}

Gravitational lensing conserves the total photon flux $F=L/(4\pi d_L^2)$, where $d_L$ is the luminosity distance. This is a fully nonlinear and general result based on the analysis of lightrays in curved spacetime. In the case of the CMB, this implies that a spatial average of the flux must be conserved from one time to the next. Because the CMB is emitted when the universe is very smooth, this implies that we should expect $\langle F\rangle\simeq \bar F$ where $\bar F$ is the value in the background. 
Since $d_L=(1+z_s)^2d_A$, this implies 
the average of $d_A^{-2}$ should be close to its background value at the same observed redshift. This means,  $\langle d^\text{eff}_A(z_s)^{-2}\rangle \simeq \bar{d}_A(z_s)^{-2}$, where the approximate equality means there are no significant corrections, i.e.,  no corrections as large as $\mathcal{O}(\delta^2)$, where $\delta$ is the matter over-density. We expand \eqref{bdjscksjdb} perturbatively,
\be
d_A^{-2}=\frac{(1+z)^2}{\chi^2}\left[1 - 2\frac{\delta d_A}{\chi}
-\frac{\delta^2 d_A}{\chi}+3\left(\frac{\delta d_A}{\chi}\right)^2\right]\,.
\ee 
The  linear term vanishes on average by definition: $\langle \delta d_A\rangle=0$. Then, 
flux conservation implies, on average, $\langle\delta^2 d_A\rangle\simeq 3 \langle(\delta d_A)^2\rangle/\chi$ and consequently
\be\label{e:Delta}
\langle\Delta\rangle\simeq \frac{3}{2}\left\langle\left(\frac{\delta d_A}{\chi}\right)^2\right\rangle=\frac{3}{2}\left\langle\kappa^2\right\rangle\,.
\ee
where $\kappa$ is the usual linear lensing convergence. This is actually the leading contribution to the expected change to large distances. We prove this remarkably simple and important result in a variety of ways in several appendices. It implies that the total area of a sphere of constant redshift will be larger than in the background. Physically this is because a sphere about us in redshift space is not a sphere in real space~-- lensing implies that this `sphere' becomes significantly crumpled in real space, and hence has a larger area. When interpreted as a shift to the background geometry, this would have important implications for the analysis of the CMB. An observed patch of the CMB sky such as a hot or cold spot  of a fixed observed angular scale will correspond to a physical area which is larger than the background value, since the distance to it is larger. 
Effectively, it is the angular size of these hot and cold spots, combined with a theoretical model for calculating both the distance to the CMB and the sound horizon scale at last scattering, that determine many key parameters of the cosmological standard model. Consequently, we anticipate a shift in the inferred background cosmology when aggregated lensing is taken into account. 

\begin{figure}[t!]
\includegraphics[width=0.95\columnwidth]{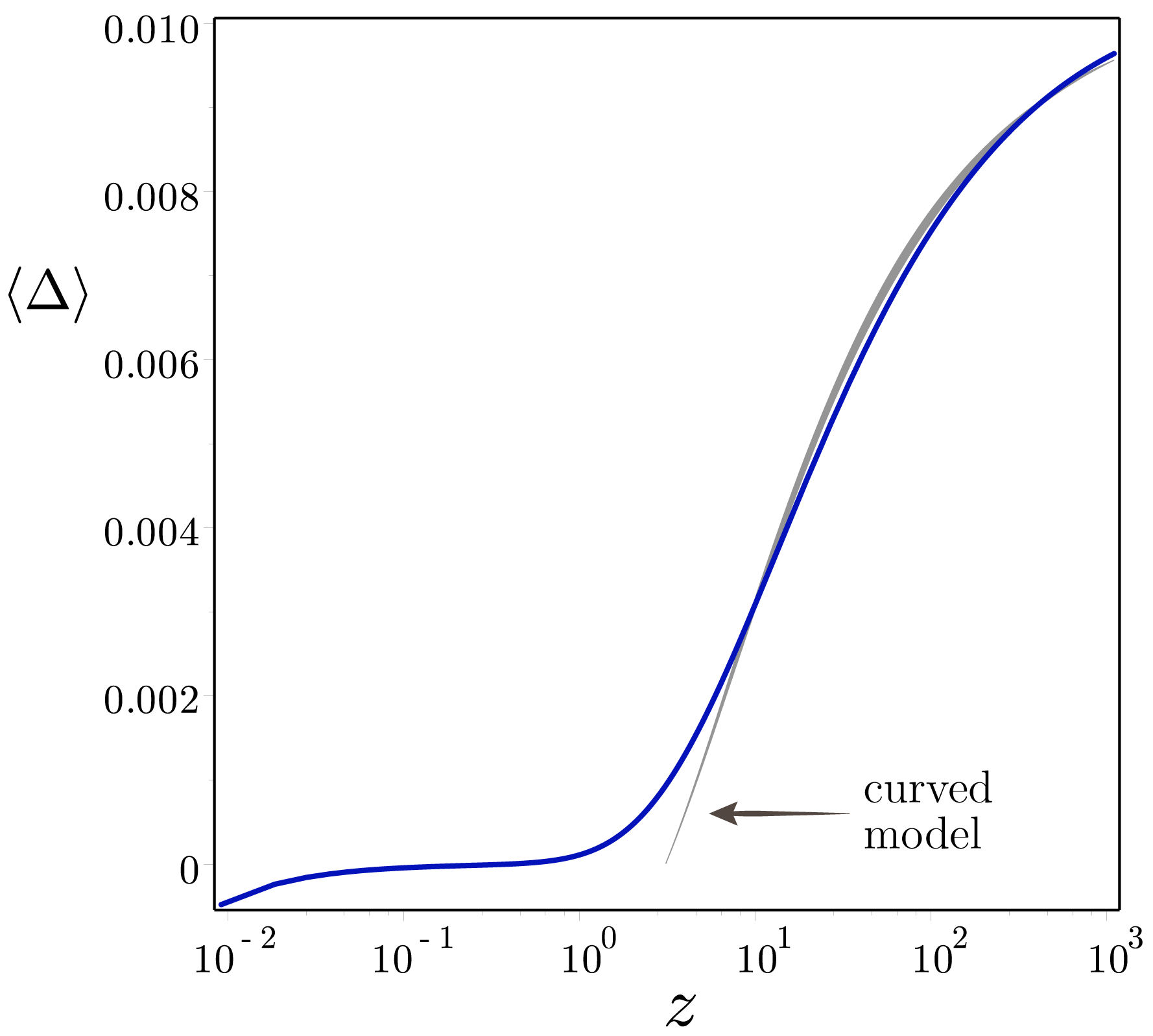}
\caption{{Fractional correction $\langle\Delta\rangle(z)$ to the distance [see \eqref{ens}]} for a fiducial model $\Omega_m=0.3,h=0.68,\omb^2=0.0222,w=-1$ and $n_s=0.96$. The correction is negative for $z\lesssim0.25$, purely from the local contribution. At higher redshift the shift arises from the aggregated lensing term~\eqref{jdshbcsjhbca}. For $z\gtrsim 10$ the corrections grow  $\propto\chi_s^3$, and are similar to an open $\Lambda$CDM model with $\Omega_K^\text{eff}\approx 0.0066$ ({grey} `curved', shown for high $z$).}
\label{sdjhbcsdb}
\end{figure}

Here we quantify this shift for a flat $\Lambda$CDM (concordance) background, see the result 
plotted in ~Fig.~\ref{sdjhbcsdb}, and we explore the potential consequences for precision cosmology.
At low redshifts the change to $\bar d_A$ is small ($|\langle\Delta\rangle| \lesssim 10^{-4}$), negative and dominated by local effects (from coupled velocity and Sachs-Wolfe terms). 
Recently \cite{Ben-Dayan:2014swa} have estimated the effect of this change on $H_0$ and especially on its variance measured with low-$z$ data. 
For $z\gtrsim 0.5$ the change becomes positive and is dominated by second-order lensing effects. It grows monotonically until last scattering, and the distance to the CMB is increased by about one percent.

If this shift is not taken into account, the distance-redshift relation for flat $\Lambda$CDM will appear to be that of a slightly open model ($\Omega_K>0$),  due to the slight shrinking of objects relative to the background. At high redshift $z\gtrsim 10$, the effective distance is similar to an open $\Lambda$CDM model.  At the CMB redshift, measurement of one value $d_A(z_*)$ requires a change in $H_0$ in the background model to account for this shift.

The incorporation of perturbations to the monopole of the local $d(z)$ relation was first considered in~\cite{Barausse:2005nf,Clarkson:2011uk} using a series expansion, where it was speculated to produce large contributions from small-scale power. (We discuss the contribution of small-scale power below.) A later study in~\cite{BenDayan:2012ct,BenDayan:2013gc} used `lightcone averaging' to analyze the monopole of the magnitude-redshift relation at low to moderate redshift.
The effect was predicted to be small, consistent with our low redshift results, 
but in contrast with  \cite{Barausse:2005nf,Clarkson:2011uk}. 
None of these works consider 
the impact on cosmological parameter estimation from CMB observations. Here we focus on high-$z$, where the aggregation of lensing events is more significant. 
We use ensemble averaging, so that specifically considering the monopole is unnecessary when the perturbations are statistically isotropic. The approach here is significantly simpler than the results obtained in the papers cited above. The details of the calculation, including four independent derivations of the main formula (\ref{e:Delta}), are presented in the Appendices.

\section{correction to the distance from geodesic deviation}

The second-order perturbation of $d_A$ about a flat $\Lambda$CDM background in the Poisson gauge has been given in~\cite{Umeh:2012pn,Umeh:2014ana} based on the Sachs equations (see also~\cite{Fanizza:2013doa,Marozzi:2014kua} for another approach). 

Assuming no anisotropic pressure and no vectors and tensors at linear order means the first-order modes are determined by a separable potential $\Phi(\eta,\bm x)=g(\eta)\Phi_{0}(\bm x)$, normalized such that  $g(\eta_0)=1$. Here $\eta$ is conformal time and $\eta_0$ denotes today.
The perturbation $\Delta$ splits into local terms evaluated at either the source or observer, local times integrated terms, and integrated terms~\cite{Umeh:2012pn}. The key local terms are not revelevant for us, and are discussed in Appendix~\ref{lsadvbasdb}.
 
The key integrated terms at second-order are those with a total of 4 screen-space derivatives which do not form part of a 2-divergence in the screen space. (A pure 2-divergence gives zero contribution to the ensemble average.) In Appendix~\ref{derivation1} we derive this result from the well known Sachs equation. The relevant term is, up to total divergence terms,
\bea
\Delta_\text{int}(z_s)
&=& \frac{3}{2}\left[\int_0^{\chi_s}\frac{\d\chi }{\chi}\left(1-\frac{\chi}{\chi_s}\right) \Delta_2\Phi(\chi)\right]^2
\label{jdshbcsjhbca}
\eea
Here $\De_2$ is the angular Laplacian on the sphere of observer directions. This second-order part is the key term which contributes when an ensemble average is taken.
Four different derivations of it 
are given in Appendix~\ref{derivation1} to \ref{A:3}.

In the full expressions given in the literature cited above, there are a variety of other terms, we have carefully checked they are sub-dominant to those which we discuss here. Any terms which form a total divergence disappear once we integrate over directions or perform an ensemble average. 

The potential can be written as 
\bea
\frac{\Phi(\chi,\bn)}{g(\eta)}&=&\int\frac{\d^3k}{(2\pi)^{3/2}}\Phi_0(\bm k)e^{i\chi\bm k\cdot\bn}
=\sum_{\ell m}\Phi^0_{\ell m}(\chi)Y^*_{\ell m}(\bn),\nonumber \\
\label{kjsbdnkjsd}
\Phi^0_{\ell m}(\chi)&=&\sqrt{\frac{2}{\pi}}i^\ell\int\d^3k j_\ell(k\chi)\Phi_0(\bm k) Y_{\ell m}(\hat{\bm k})\,.
\eea
The power spectrum of $\Phi_0$ is
\be
\langle \Phi_0(\bm k_1)\Phi_0(\bm k_2)\rangle=\frac{2\pi^2}{k^{3}}\mathcal{P}_0(k)T(k)^2\delta^3(\bm k_1+\bm k_2),
\ee
where $T(k)$ is the transfer function and
\be
\mathcal{P}_0=\left(\frac{3\Delta_\mathcal{R}(k_0)}{5g_\infty}\right)^2\left(\frac{k}{k_0}\right)^{n_s-1}\,.
\ee  
Here
$$g_\infty\approx \frac{1}{5}(3+2\Omega_m^{-0.45}), \quad
\Delta^2_\mathcal{R}\approx2.4\times10^{-9}$$ 
is the squared amplitude of the primordial curvature perturbation at the pivot scale 
$k_0=0.002\,\text{Mpc}^{-1}$, and $n_s-1$ is the spectral tilt.

The expectation value of the dominant integrated contribution can be reduced to a form convenient for numerical integration (see the Appendix~\ref{A:5} for details):
\bea\label{ljshdbjasdhb}
&&\langle\Delta_\text{int}\rangle= 6\pi\sum_{\ell=0}^\infty\left[\frac{\ell(\ell+1)}{2\ell+1}\right]^2 \times
\nonumber\\&&~~
\int _{0}^{\chi_{{s}}}\!{\d\chi}
\frac{(\chi_s-\chi)^2}{\chi\chi_s^2}
g^2(\chi)\big(\mathcal{P}_0T^2\big)\Big|_{k=(\ell+1/2)/\chi}.
\eea

For large distances we can estimate analytically  the scaling behaviour {when baryons are neglected}. A crude estimate of the transfer function is (adapted from~\cite{Eisenstein:1997ik})
$T(k=\ell/\chi)\approx
1/\left[1+\alpha(\ell/\chi k_\text{eq})^2\right]$, where $k_\text{eq}\approx0.075\Omega_mh^2$\,Mpc$^{-1}$ is the equality scale, and $\alpha\sim0.05$. The main contribution to the sum {in \eqref{ljshdbjasdhb}}  comes from small scales with $\ell>\chi_sk_\text{eq}$. Approximately, the~$\ell$ factors out of the integral and the sum from $\chi_sk_\text{eq}$ to infinity gives a factor of $\sim 1/(4 k_\text{eq}\chi_s)$, assuming $g\sim g_\infty$. The integral becomes $\sim (k_\text{eq} \chi_s)^4$, giving $(k_\text{eq} \chi_s)^3$ scaling. In fact, for this transfer function the sum over $\ell$ and the integral can be done analytically (ignoring the $+1/2$ in the Limber approximation), from which we find for large $\chi_s$,
\be\label{jksdbnjksd}
\langle\Delta_\text{int}\rangle \sim 2 \Delta_\mathcal{R}^2 (k_\text{eq} \chi_s)^3\approx 0.014\left(\frac{\Omega_mh^2}{0.14}\right)^3\left(\frac{\chi_s}{14\,\text{Gpc}}\right)^3.
\ee
(This estimate is a reasonable approximation to the numerical result at large distances~-- but note it is very sensitive to $\alpha$).
For a standard cosmology this implies corrections around the percent level for 10\,Gpc distances, making {$\langle\Delta_\text{int}\rangle$} the dominant part of the signal for $z\gtrsim1$.

The generic behaviour of $\langle\Delta\rangle(z)$ is shown in Fig.~\ref{sdjhbcsdb}: at low $z$ the local Doppler contribution dominates, and the amplitude is small, $\mathcal{O}(10^{-4})$, and negative for $z\lesssim0.2-0.3$ depending on the model. At higher $z$, the amplitude is positive, implying larger distances, and grows roughly linearly in $z$ reaching near percent-level around {$z\sim5-10$}, thereafter growing proportional to the volume [roughly approximated by \eqref{jksdbnjksd}] reaching around 1\% by $z\sim10^3$. This is the aggregated lensing signal. A higher matter density or Hubble constant increases the amplitude of $\langle\Delta\rangle$, while increasing the baryon fraction or including a tilt to $n_s<1$ decreases it by tens of percent.

\subsection*{Small-scale sensitivity}

The convergence of the sum in~\eqref{ljshdbjasdhb} is very slow, reflecting the sensitivity to the accumulation of many small-scale lensing events. 
We can attempt to estimate the convergence rate analytically to determine the modes that are important. For the transfer function and approximations leading to \eqref{jksdbnjksd}, we may replace the formal sum to infinity with a cutoff at $\ell_\text{max}=\chi_s k_\text{max}$, implying that the sum now gives a contribution $(1/k_\text{eq}-1/k_\text{max})/4\chi_s$. Consequently for 
percent-level accuracy we need $k_\text{max}\gtrsim100k_\text{eq}\sim1$\,Mpc$^{-1}$. This seems reasonable, but actually overestimates the convergence rate significantly. Including the logarithmic amplification of small-scale power in a more accurate transfer function slows the convergence so that percent accuracy is achieved for only for $k_\text{max}\gtrsim10^4k_\text{eq}$, while $k_\text{max}\sim100k_\text{eq}$ is only 10\% accurate. This slow convergence  remains also  when baryons are included or {small-scale} power is amplified by a factor of 2 according to~\cite{Smith:2002dz}.  \emph{This implies that modes down to 10kpc scales in principle contribute to the aggregated lensing effect, and cannot be neglected.}  
Since the power spectrum is not precisely modelled on these scales, we present our results for the 
 linear,  analytical approximation to the power spectrum given in 
Ref.~\cite{Eisenstein:1997ik}. This in general underestimates the true power spectrum on small scales,  however, most of the effect accumulates at high $z$ where linear theory is accurate, so our key results should be correct.

One may ask whether we should add an ultraviolet cutoff to smooth away small scale power in $\Phi$. This is appropriate when binning data and calculating correlation functions, for example, as the data sample effectively smooths over the linear matter distribution. This is not the case for the cumulative relativistic correction considered here, which is effectively a sum over all possible lensing events to give an ensemble average. Note also that, neglecting the mild time dependence of $\Phi$ the expression in \eqref{jdshbcsjhbca} is positive definite and smoothing does nor reduce it.

The physically relevant dark matter free-streaming cutoff is $\mathcal{O}(\text{pc})$, but this does not change our results.  A larger UV-cutoff~-- equivalent to smoothing $\Phi$~-- simply investigates a different ensemble of universes with different lensing properties on small scales, quite unlike the real universe, and underestimats the size of the effect (linear theory with no cutoff still slightly underestimates it, but not by much). For small objects like SNIa it is clear we need high $k$-modes as the {lightrays probe} such small scales~\cite{Clarkson:2011br}. For a diffuse background like the CMB it is tempting to assume that the CMB light effectively smooths the matter distribution on the scale of observation. That is true at linear order, but not at second order, where the lensing effects accumulate and do not cancel in the same way. 
Mathematically, if we smooth $d_A(z,\bn)$ over a solid angle on the sky, we would eventually integrate  \eqref{sakjncsdacjkn} against a window function in $\bn$; but this integral simply drops out after applying the spherical harmonic addition theorem, making no difference to $\langle\Delta\rangle$. This is not the same as a cutoff in $k$, which smooths $\Phi$ itself, not $d_A(z,\bn)$. Consequently, it is important not to smooth the potential, and include power down to the smallest relevant scales. 
\begin{figure*}[ht!]
\includegraphics[width=0.9\textwidth]{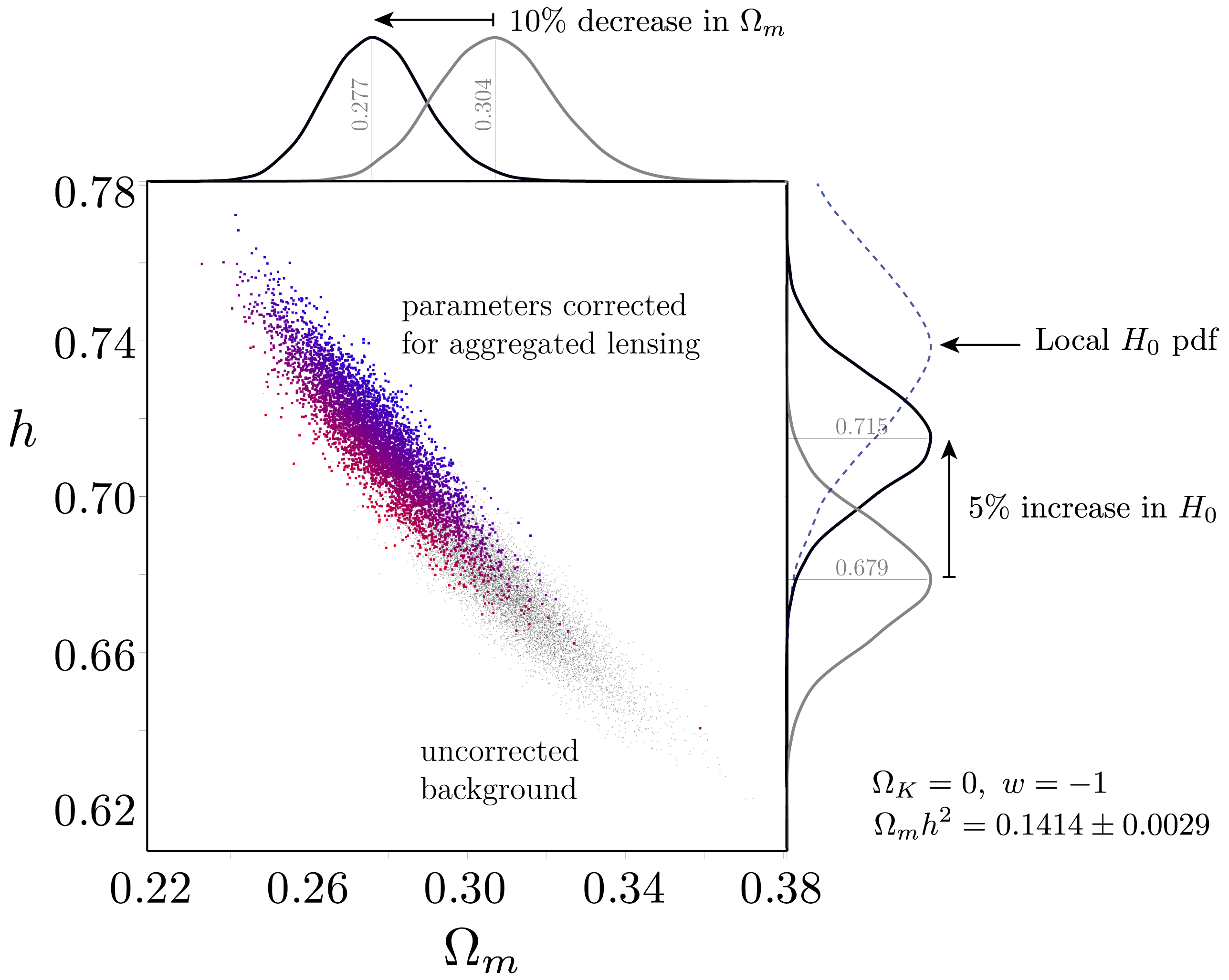}
\caption{ \label{fig:hcorr} We show the distribution of values for $h$ and $\Om_m$
from the Planck data when using only the background angular diameter distance (black dots) and when including the correction $\Delta_*$  (red to blue dots, coloured according to $d_A(z_*)$, with red indicating larger distances).   At the boundary we also show the one dimensional probability distribution function of the corresponding variable. The resulting Hubble parameter is now in agreement with local measurements~-- the dotted PDF is adapted from Riess et al.~\cite{Riess}.
}
\end{figure*}

\section{potential Implications for the CMB and cosmological parameter estimation}

We now explore the potential effects on parameter estimation from the CMB under the assumption that the correction from aggregated lensing can be interpreted as a monopole shift in the distance. This assumption requires further justification, as we discuss in the Conclusions, and it is likely that our parameter estimates will be shifted when the full effects of aggregated lensing are taken into account. We leave this for further work.

If the distance to the CMB is changed by a homogeneous factor $(1+ \Delta_*)$,  $d_A \rightarrow (1+ \Delta_*)d_A$, features in the CMB are simply rescaled. More precisely, within the flat sky approximation which can be used for $\ell>20$, the CMB power spectrum $C_\ell$ changes as~\cite{Vonlanthen:2010cd}
\be
C_\ell  \rightarrow (1+ \Delta_*)^2C_{(1+ \Delta_*)\ell}\,.  
\ee
We find that the shift in the distance to the CMB is well approximated by 
 \be
 \Delta_*=0.63\, \Omega_m^{1.1f_b+1.8} h^{2.5f_b+3} n_s^5 e^{-0.28\Om_b/0.0222}
 \ee
where the baryon fraction is $f_b=\Om_b/\Om_m$. This is accurate to about 1\% in the vicinity of the Planck parameter values, and 10\% accurate well outside that range. 

Using the Planck parameters for $\Om_mh^2\equiv\omm$, $h$, $\Om_b h^2\equiv\omb$ and $n_s$ (as independent Gaussian distributed values from `Planck+Lensing', Table 2 in~\cite{Ade:2013zuv}, and setting $z_*\approx1090$) we find 
\be
\langle\Delta\rangle(z_*)\equiv \Delta_*= 0.010285\pm 0.00074\,.
\ee 
Hence aggregated lensing leads to a change of about 1\% in the distance to the CMB which is extremely important for CMB parameter estimation. This corresponds to  $\sim$2.5$\sigma$ error on the Planck estimation of the distance, so we anticipate that background parameters dependent on this distance shift by about 2.5$\sigma$. (Of course the Planck parameters are not independent but to some extent correlated, but this does not modify the mean value of $\Delta_*$, only its error bar.) The CMB measures $d_A(z_*)$ precisely, so we must adjust $\bar d_A(z_*)$ to account for $\Delta_*$. 

In Fig.~\ref{fig:hcorr} we compare the values of $h$ and $\Om_m$ inferred without correction with those obtained including aggregated lensing. For this we use the Planck values~\cite{Ade:2013zuv}  $\omm=0.1414\pm 0.0029$, and $h=0.679 \pm0.015$ giving the measured distance $(1+z_*)d_A(z_*)=13.98\pm 0.13$Gpc (we ignore the small correction from radiation), and we take $\omb =0.02217\pm 0.00033$, $n_s=0.9635\pm 0.0094$.  We treat these as independent Gaussian distributed variables. The values of $\omm$, $\omb$ and $n_s$ are determined by the physics of last scattering only, and do not depend on the distance~-- but $h$ does. 
From $d_A(z_*)=\bar d_A(z_*)(1+\Delta_*)$ we then infer the background parameters $h$, 
$\Om_m$. For a flat $\La$CDM model we find:
\be
h=0.715\pm0.0153,~~~~\Omega_m=0.277\pm0.0125
\ee
For $h$ this is now in good agreement with  Riess et al.~\cite{Riess},  $h=h_R= 0.738\pm 0.024$, removing all tension. 

At first one might be surprised that a 1\% shift in $d_A$ yields such a large (about 5\%) shift in $h$ and $\Omega_m$. However, a short calculation shows that this is to be expected: at fixed $\omm$, neglecting curvature and radiation we have
\bea
d_A(z_*) &=& \frac{L_0}{1+z_*}\int_0^{z_*}\frac{dz}{[\omm(1+z)^3-\omm +h^2]^{1/2}}  \nonumber\\
\frac{\dd d_A}{\dd h}(z_*) &=& \frac{-hL_0}{1+z_*}\int_0^{z_*}\frac{dz}{[\omm(1+z)^3-\omm +h^2]^{3/2}}\,. \nonumber
\eea
where $L_0=100$km\,s$^{-1}$Mpc$^{-1}\simeq 3000$Mpc. If we want to absorb a change in $d_A$ as a change in the background value of $h$ we use the fact that the background value of the distance needs to be reduced by $\Delta$ from its observed value: 
\bea
d_A^\text{obs}&=&\bar d_A(\bar h)(1+\De)~~~\text{implies}\\
\bar d_A(\bar h)&=&d_A^\text{obs}(1-\De)\\
 &=& \bar d_A(h) +\frac{\dd d_A}{\dd h}h\De_h
 \eea
 we require a change in $h$, $h= \bar h(1+\De_h)$ given by 
\be
  \De_h = -\frac{d_A}{(\dd d_A/\dd h)h}\De  \simeq + 5\De \,.
\ee
For the last $\simeq$ sign we have set $z_*=1090$, $\omm=0.14$ and $h=0.72$.  Hence $h$ is very sensitive to $d_A$. A 1\% change in $d_A$ does necessitate a 5\% increase in  $h$, and consequently a 10\% decrease in $\Omega_m$, as we have observed in our analysis.

In a full joint analysis not only $h$ and $\Omega_m$ but also other parameters will shift. As an example we consider curvature~\footnote{While it is  strictly inaccurate to modify the background with curvature given our calculation of $\langle\Delta\rangle$, the errors will be $\mathcal{O}(\Omega_K\Delta)$ which is small.} or evolving dark energy with a constant equation of state, using $\omm$, $n_s$ and $d_A(z_*)$ as determined by Planck together with 
the Riess et al. value of the Hubble parameter. This yields shifts in the background parameters by about 1$\sigma$:
\bea
100\,\Om_K &=& +0.37\pm 0.47,  \qquad (\text{for}~~w=-1)\\
w&=& -1.07\pm 0.082,  \quad~ (\text{for}~~\Om_K=0)\,.
\eea
Hence taking this shift to the distance into account together with local observations, the CMB data remains consistent with a minimal flat $\La$CDM model. This is in contrast to the standard calculation of distance where it is difficult to relieve the tension between local measurements of $H_0$~\cite{Ade:2013zuv,Riess:2011yx,Marra:2013rba,Busti:2014dua} and the value from CMB observations.

Clearly, the analysis presented here is not definitive for several reasons. First, as mentioned above, the Planck measurements of cosmological parameters are not independent and especially the Planck value of $\omm$ is not completely independent of the distance $d_A$. We have also assumed a simple linear transfer function~\cite{Eisenstein:1997ik}. A full likelihood analysis should be performed with $\langle\Delta\rangle$ properly included.

Furthermore, aggregated lensing not only leads to a average shift in the distance to the CMB but $\Delta$ is actually direction dependent~\cite{Umeh:2012pn}. We expect its fluctuations to be imprinted as additional
fluctuations in the CMB. However, since the dominant contribution comes from very small
scales, we expect them to show up mainly at  high $\ell
> 2000$ and we believe that the effect on the mean distance discussed here is the dominant one in present CMB experiments.
To do a fully consistent analysis, which combines second order fluctuations in the distance  with temperature perturbations, a 3rd order Boltzmann solver would be needed. A interesting future project which is (far!) beyond the scope of the present paper.

We also note that higher-order contributions to $\langle\Delta\rangle$ will be small, though ultimately necessary as observations improve. We anticipate they will be dominated by terms such as $(\delta d_A)^4$, which will lead to a percent level correction to our second-order correction. Thus, the main contribution to aggregated lensing is from~\eqref{jdshbcsjhbca}.

\section{Conclusions}

We have demonstrated an important overall shift in the distance redshift relation when the aggregate of all lensing events is considered,  calculated by averaging over an ensemble of universes. This result is a consequence of flux conservation at second-order in perturbation theory. This is a purely relativistic effect with no Newtonian counterpart~-- and it is the first quantitative prediction for a significant change to the background cosmology when averaging over structure~\cite{Clarkson:2011zq}. The extraordinary amplification of aggregated lensing comes mainly from the integrated lensing of structure on scales in the range $1-100$\,Mpc, although structure down to 10kpc scales contributes significantly. We have estimated the size of the effect using a linear transfer function which slightly underestimates power on small scales at high redshift, so this provides a robust lower limit to the overall amplitude. Higher-order corrections from relativistic perturbation theory will enter $\mathcal{O}((\delta d_A)^4)$, making \eqref{jdshbcsjhbca} the main contribution in general. 

This isotropic shift is particularly important for high redshift, apparently giving a change to the distance to the CMB of one percent. What does this mean?  We have argued that the shift can be interpreted as a change to the inferred background cosmology. Assuming that observations of the CMB really measures the area distance implies that fitting to the minimal $\La$CDM model  leads to an underestimation of the Hubble parameter by 5\%.
We have considered the consequences for analysing the CMB, and have argued that parameter estimation could be strongly affected~-- parameter constraints can be shifted by more than $1\sigma$. Because the shift $\langle\Delta\rangle$  increases the distance relative to the background, the corrections to the background to compensate for this increase in  distance are achieved by increasing $h$. In particular, we have shown that a higher $h$ is naturally preferred over the low value found by Planck~\cite{Ade:2013zuv}, in line with local data~\cite{Riess:2011yx}.  For current and future redshift surveys, the effective model yields sub-percent changes to the curvature, and dark energy parameters, which will eventually be important for precision cosmology of the coming decade. 

Of course, our analysis is rather simplistic. We have calculated the expected correction to the distance of an infinitesimal spot on the CMB sky and extrapolated up to sound-horizon scales. We have assumed that the CMB temperature map directly measures the distance to last scattering, when in fact it is inferred indirectly. The full correction to the background parameters corresponding
to aggregated lensing~-- a change to the global distance
from the sum of all lensing events~--  may well give quite
different results. Nevertheless, as our simple consideration has led to 5\% difference in the estimated Hubble parameter, and 10\% in $\Omega_m$, it is clear that 2nd order aggregated lensing has to be included in a CMB calculation which aspires to a precision of 0.1\%.
Beyond linear perturbation theory, the correspondence between the expectation value of observables and the averaged background model becomes very subtle to define precisely. We have shown that it could lead to important corrections to the background when asking, What is the distance to the~CMB?

~\\[5mm]

{\bf Acknowledgements:} 
Special thanks to Camille Bonvin, Antony Lewis and Fabien Schmidt for extensive and very helpful discussions. We are also grateful for discussions with and comments from Phil Bull, George Ellis, Pedro Ferreira, Daniel Holz, Martin Kunz, Giovanni Marozzi, Uros Seljak, Bjoern Schaefer, Bob Wald and David Weinberg.
CC and RM are supported by the South African National Research Foundation. OU and RM  are supported by the South African Square Kilometre
Array Project. RM acknowledges support from the UK Science \& Technology Facilities Council (grant ST/K0090X/1). RD acknowledges support from the Swiss National Science Foundation.

\appendix
\begin{widetext}

\section{The derivation of \eqref{jdshbcsjhbca} geodesic deviation}\label{derivation1}

The dominant terms in the ensemble average come from those with the largest number of screen space derivatives in them. Since a pure divergence at second-order will give zero ensemble average (or monopole), we can present a simple derivation of the relevant term.

We consider a bundle of future-pointing light rays around a given ray with momentum $k^a$, $k^b\nabla_b k^a =0$, $k_a k^a=0$. A light ray arriving at the observer from a slightly different direction is given by its  angular position at the observer, $\bm\alpha$. The Jacobi map maps $\bm\alpha$, to a screen space position $\bm x_\perp(\la)$, where $\la$ is the affine parameter along the geodesic $k^a$.
As $\d  x_\perp^a/\d\la =  x_\perp^b\nabla_b k_a$ (see e.g.~\cite{SEF}, eq. (3.50)), the area on the screen space satisfies
\be
 \frac{\d{\cal A}}{\d\la} = \nabla_a k^a{\cal A} = \theta{\cal A} \,.
 \ee
Where $\theta =\nabla_a k^a$ is the rate of expansion of the null congruence. Defining the area distance
\be d_A \propto \sqrt{\cal A}\,, \ee
we obtain the basic equation for the area distance, valid in any spacetime:
\be\label{dslkjcbskjc}
k^a\nabla_a d_A=\frac{1}{2}\theta d_A \,.
\ee
This definition gives the proper area of a source, $\d {\cal A}$, in terms of the solid angle at the observer, $\d\Omega$, as $\d {\cal A}= d_A^2 \d\Omega$. We expand order-by-order on a perturbed Minkowski background where $\bar d_A=\chi=\eta_0-\eta$, $\bar\theta=-2/\chi$, $\bar k^a\nabla_a=-D/\d\chi$, as $d_A=\bar d_A+\delta d_A+\frac{1}{2}\delta^2 d_A$, similarly $k^a + \delta k^a  + \frac{1}{2}\delta^2 k^a$ and  $\theta=\bar\theta+ \delta \theta  + \frac{1}{2}\delta^2\theta$. Photon geodesics are conformally invariant, hence we can remove the irrelevant cosmic scale factor in this calculation. A background conformal transformation then gives the full area distance~\cite{Umeh:2014ana}. 

 The first order perturbation of the area distance is well known, see, e.g. ~\cite{Umeh:2014ana,Bonvin:2005ps}
\bea
\delta d_A(\chi)&=&\int_0^\chi\d\tilde\chi (\tilde\chi-\chi)\tilde\chi \nabla_\o^2\Phi(\tilde\chi)\,.
\eea
Inserting this in~\eqref{dslkjcbskjc} and neglecting subdominant terms (at first-order the perturbation of $\delta k^a$ contains at most one $\nabla_\o$), we obtain the perturbation of the null expansion 
\bea\label{e:theta1}
\delta\theta(\chi)&=& 
-2\frac{\d}{\d\chi}\frac{\delta d_A}{\chi}
= 2\int_0^\chi\d\tilde\chi \left(\frac{\tilde\chi}{\chi}\right)^2 \nabla_\o^2\Phi(\tilde\chi)\,\label{dsjhbcsdhjb}
\eea
At second-order,
\be
-\frac{\d}{\d\chi}\delta^2d_A+2\delta k^a\nabla_a\delta d_A+\delta^2k^a\nabla_a\chi = \frac{1}{2}\chi\delta^2\theta +\delta d_A\delta\theta+\frac{1}{2}\bar\theta \delta^2 d_A
\ee
The left hand side of this equation contains the relevant post-Born-approximation terms which reflect the fact that the geodesic deviates from a straight line. The right hand side is the impact of the change of the rate of expansion of the geodesic congruence on the distance. 
Now, we can neglect: $\delta^2k^a\nabla_a\chi=\delta^2k\p\nabla\p\chi$ as it has only lower derivative terms. Then $\delta^2\theta\sim\nabla_{\o i}\delta^2 k^i+$lower derivative terms. Since the leading order is a 2-divergence it will not contribute to the mean.\footnote{We can see this once we have expanded in spherical harmonics we find  $
\sum_{m=-\ell}^{m=\ell} Y_{\ell m}(\bm n)Y^*_{\ell m}(\bm n)=({2\ell+1})/{4\pi}
$ implies $\sum_{m=-\ell}^{m=\ell} \nabla_{\Omega}^iY_{\ell m}(\bm n)Y^*_{\ell m}(\bm n)=0$, and the divergence of this is zero too. }
This leaves
\bea
\frac{\d\delta^2d_A}{\d\chi}-\frac{1}{\chi} \delta^2 d_A&=& -\delta d_A\delta\theta+2\delta k^a\nabla_a\delta d_A \\
&=&-3\delta d_A\delta\theta = 6 \delta d_A\frac{\d}{\d\chi}\frac{\delta d_A}{\chi}\label{kjdsnvsjkdn}
\eea
where the second equality holds up to a pure divergence on writing $\nabla_{\o i}(\delta k^i\delta d_A)=\nabla_{\o i}(\delta k^i)\delta d_A+\delta k^i\nabla_{\o i}(\delta d_A)$ and we have used (\ref{e:theta1}) for the last equality. 
Noting that the right hand side of~\eqref{kjdsnvsjkdn} equals  $\chi\frac{\d}{\d \chi}(\frac{\delta^2d_A}{\chi})$ we obtain
\bea \label{e:A9}
\frac{\delta^2 d_A(\chi_s)}{\chi_s} &=&{3}\left(\frac{\delta d_A(\chi_s)}{\chi_s}\right)^2
\\
\delta^2 d_A(\chi_s)&=& {3}{\chi_s}\left[\int_0^{\chi_s}\d\chi \left(1-\frac{\chi}{\chi_s}\right)\chi \nabla_\perp^2\Phi(\chi)\right]^2\\  &=& 3\chi_s
\left[\int_0^{\chi_s}\frac{\d\chi}{\chi} \left(1-\frac{\chi}{\chi_s}\right) \De_2\Phi(\chi)\right]^2
\label{e:A8}
\eea
For the last equal sign we used $\chi^2\nabla_\perp^2 =\De_2$.

 This can be interpreted in a variety of different ways, on moving the angular derivatives outside the integrals, and expanding the pure divergence $\nabla_{\o i}[\nabla_{\o j}(\nabla^{\o i}X \nabla^{\o j}Y)]$. This can be used to re-write this in terms of the (linear null shear)$^2$ and the bending angle coupled to gradients in the linear convergence as presented in~\cite{Umeh:2012pn}.

\subsection*{Consistency with the Sachs equation}

For completeness, we can show consistency with the Sachs equations, written in terms of the distance. We use the notation of Ref.~\cite{Umeh:2012pn}.
\begin{equation}\label{CovAreadistance}
\frac{\d^2 d_A}{\d\lambda^2}=-\frac{1}{2}\left[  R_{ab} k^a k^b+{\Sigma_{ab}}\Sigma^{ab}\right] {D}_A
\end{equation}
where the shear obeys
\begin{eqnarray}\label{eq:shearevo}
\frac{\text{D}{\Sigma}_{\<ab\>}}{\d\lambda}&=&-\theta{\Sigma}_{ab}+ N_{\<a}{}^{e} N_{b\>}{}^{f}{R}_{efcd} k^c k^d \,.
\end{eqnarray}
These are fully non-linear equations valid in any spacetime. 
The second-order part of \eqref{CovAreadistance} is
  \begin{eqnarray}
  \frac{\d^2\delta^2 d_A}{\d\lambda^2}&=& -2\delta k^a\delta k^b \nabla_a\nabla_b d_A - 2k^b \delta^2 k^a \nabla_a \nabla_b d_A-4 \delta k^ak^b \nabla_a\nabla_b \delta d_A - 2 \delta k^a \nabla_a \delta k_b\nabla^b d_A \nonumber\\ &&
 -{\text{D}\delta^2 k_a\over \d\lambda} \nabla^a d_A- \delta^2k^a \nabla_a k_b\nabla^b d_A- 2 {\text{D}\delta k_a\over \d\lambda} \nabla^a \delta d_A - 2 \delta k^a \nabla_a k_b \nabla^b \delta d_A 
 -\delta d_A\delta R_{ab}k^ak^b
  \nonumber\\ &&
  +d_A\left[- \delta \Sigma_{ab}\delta \Sigma^{ab}-2\delta k^ak^b\delta R_{ab} -\frac{1}{2} \delta^2 R_{ab}k^ak^b \right] \,. \label{eq:areadistsecond1}
\end{eqnarray}
Neglecting contributions that do not have 4 derivatives this becomes
  \begin{eqnarray}
  \frac{\d^2\delta^2 d_A}{\d\chi^2}&=& 
  -4 \delta k^ak^b \nabla_a\nabla_b \delta d_A 
 - 2 {k^b\nabla_b\delta k_a} \nabla^a \delta d_A 
 - 2 \delta k^a \nabla_a k_b \nabla^b \delta d_A 
 -\delta d_A\delta R_{ab}k^ak^b
  -\chi \delta \Sigma_{ab}\delta \Sigma^{ab}
 \,. \label{eq:areadistsecond1}
\end{eqnarray}
Next we can calculate, including only 4-derivative terms that do not form a divergence [$'=d/d\chi$]:
\bea
\delta k^a k^b \nabla_a\nabla_b \delta d_A&=&\delta\theta\delta d_A'\\
{k^b\nabla_b\delta k_a} \nabla^a \delta d_A&=& \frac{1}{\chi}\delta\theta\delta d_A +\delta\theta'\delta d_A\\
\delta k^a \nabla_a k_b \nabla^b \delta d_A&=& \frac{1}{\chi}\delta\theta\delta d_A\\
\delta d_A\delta R_{ab}k^ak^b&=&-2\nabla_\o^2\Phi\delta d_A\,.
\eea
For the shear we solve \eqref{eq:shearevo}:
\be
\delta\Sigma_{ij}= 2\int_0^\chi\d\tilde\chi \left(\frac{\tilde\chi}{\chi}\right)^2 \nabla_{\o\langle i}\nabla_{\o j\rangle }\Phi(\tilde\chi)\,.
\ee
Now, noting that, up to a pure divergence,
\be
\nabla_{\o\langle i}\nabla_{\o j\rangle } A \nabla^{\o\langle i}\nabla^{\o j\rangle }B=\frac{1}{2}\nabla_\o^2A\nabla_\o^2B,
\ee
we find
\be
\delta\Sigma_{ij} \delta\Sigma^{ij}=\frac{1}{2}(\delta\theta^2) + \mbox{ divergence.}\
\ee
Inserting these results in (\ref{eq:areadistsecond1}) we find
\be
 \frac{\d^2\delta^2 d_A}{\d\chi^2}=-4\left(\frac{\d}{\d\chi}+\frac{1}{\chi}\right)\delta\theta\delta d_A+2\nabla_\o^2\Phi\delta d_A-\frac{1}{2}\chi\delta\theta^2\,.
\ee
Next we use
\bea
\frac{\d}{\d\chi}\delta\theta&=&2\nabla_\o^2\Phi-\frac{2}{\chi}\delta\theta\,,\\
\frac{\d}{\d\chi}\delta d_A &=& \frac{1}{\chi}\delta d_A-\frac{1}{2}\chi\delta\theta\,;
\eea
these imply
\be
2\nabla_\o^2\Phi\delta d_A-\frac{1}{2}\chi\delta\theta^2 = \left(\frac{\d}{\d\chi}+\frac{1}{\chi}\right)\delta\theta\delta d_A
\ee
so that
\be
 \frac{\d^2\delta^2 d_A}{\d\chi^2}=-3\left(\frac{\d}{\d\chi}+\frac{1}{\chi}\right)\delta\theta\delta d_A\,,
\ee
which is the derivative of 
\be
\frac{\d\delta^2d_A}{\d\chi}-\frac{1}{\chi} \delta^2 d_A
=-3\delta d_A\delta\theta\,,
\ee
in agreement with \eqref{e:A8}.

\section{Integral formulation, and the meaning of linear terms at second-order}\label{A:3}

Here we present a more conventional integral formulation of the derivation of \eqref{jdshbcsjhbca}.  
We integrate~\eqref{dslkjcbskjc} directly along the perturbed geodesic:
\bea
2\ln d_A(\chi_s,\bm n_\text{obs}) &=& \int_{\lambda_o}^{\lambda_s} \d\lambda\theta \\
&=&-\int_{\lambda_o}^{\lambda_s} \d\lambda\frac{2}{\chi}+\int_{\lambda_o}^{\lambda_s} \d\lambda\delta \theta +\int_{\lambda_o}^{\lambda_s} \d\lambda\delta^2\theta\label{sdajhb}
\eea
Note that the left hand side is a scalar evaluated at the \emph{observed} source position $S$, as we integrate from observer to source. On the right hand side, we have expanded $\theta$, and have kept the integrals along the perturbed geodesic given by the curve $x^a(\lambda)=\int_{\lambda_o}^{\lambda}\d\tilde\lambda k^a(\tilde\lambda)$.
The first term of \eqref{sdajhb} is trivial to evaluate, and the last leads to a 2-divergence. The middle one, which is the integral of a first order quantity on a perturbed path which we need to second-order. Referring to Fig.~\ref{fig:Chris}, we can expand the integral about a background geodesic in different ways:\\ 
\begin{figure}[ht!]
\includegraphics[width=0.9\textwidth]{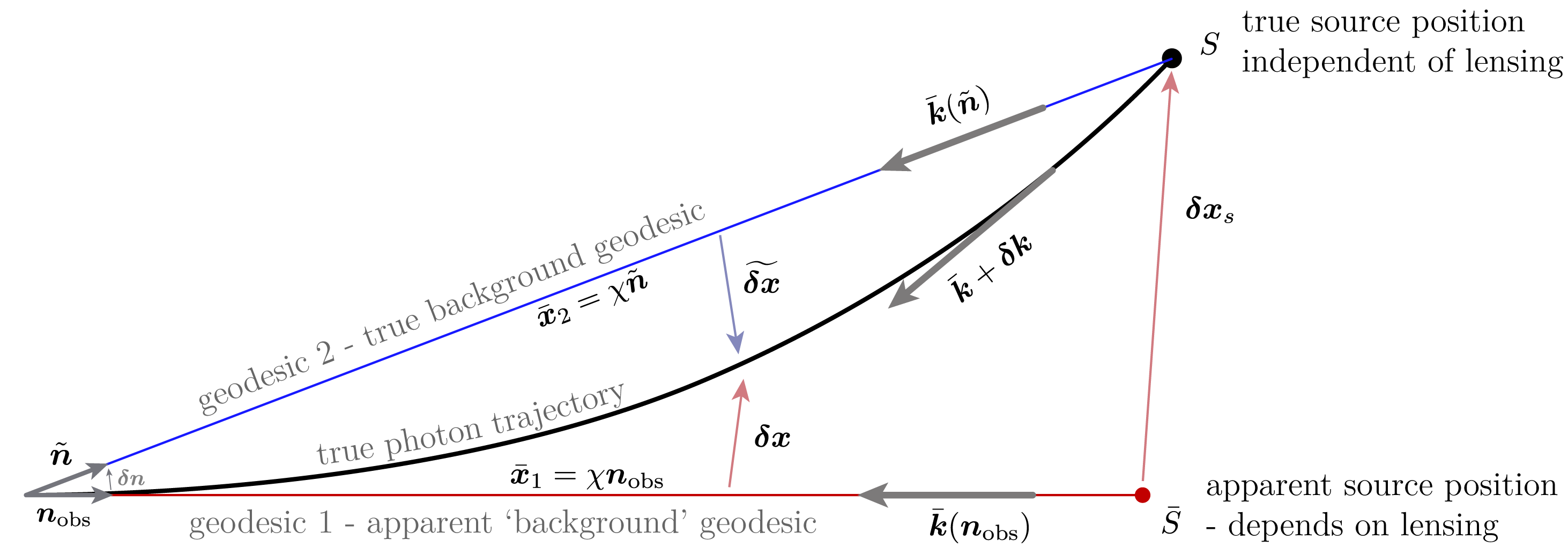}
\caption{\label{fig:Chris} Real and apparent source positions relative to the observer give different `background' geodesics: \\
{\bf geodesic 1:} the geodesic from the observer in the same physical direction as the observed source, $\bm n_\text{obs}$. This straight line ends at a fictitious place for the source - the apparent position.\\
{\bf geodesic 2:} the geodesic from the observer to the physical position to the  observed source. This straight line starts at an unobservable place for the observer.}
\end{figure}
Let us consider expanding about geodesic 1. 
\bea
\int_{\lambda_o}^{\lambda_s} \d\lambda\delta \theta(\bm n_\text{obs},x(\lambda))&=&
\int_{\lambda_o}^{\lambda_s} \d\lambda\delta \theta(\bm n_\text{obs},\bar x_1(\lambda))+\int_{\lambda_o}^{\lambda_s} \d\lambda\delta x^i\nabla_i\delta \theta(\bm n_\text{obs},\bar x(\lambda))\label{dhjsb}
\eea
We can write on any \emph{background} geodesic
\be
\int_{\lambda_o}^{\lambda_s} \d\lambda\delta \theta(\bm n,\bar x(\lambda)) =-\int_{0}^{\chi_s} \d\chi\delta \theta(\bm n,\bar x(\chi)) = 2\frac{\delta d_A}{\chi} \,.
\ee
 On the right hand side of this, we must specify the location the function is evaluated at, either in terms of a physical spacetime position, or a direction from the observer. 

 Then we can write~\eqref{dhjsb} as:
\bea
\int_{\lambda_o}^{\lambda_s} \d\lambda\delta \theta(\bm n_\text{obs},x(\lambda))&=&
2\frac{\delta d_A}{\chi}\bigg|_{\bar x_{\tilde{s}}}
+\int_{\lambda_o}^{\lambda_s} \d\lambda\delta x^i\nabla_i\delta \theta(\bm n_\text{obs},\bar x(\lambda))\\
&=&
2\frac{\delta d_A}{\chi}\bigg|_{ x_{{s}}}
-2\delta x^i\nabla_i\frac{\delta D}{\chi}
+\int_{\lambda_o}^{\lambda_s} \d\lambda\delta x^i\nabla_i\delta \theta(\bm n_\text{obs},\bar x(\lambda))
\eea
where the second line shifts the linear term evaluated at the image position $\bar S$ to the true source position $S$. In the second line we now have a result where both sides are evaluated at the same physical point, whereas the first line is an equation where the left hand side is evaluated at the true source position $S$, and the right hand side at the apparent position $\bar S$. If we take an ensemble average at this point, the first term on the second line is linear, evaluated at the true source position, and can be ignored. 

Up to a total divergence we then obtain 
\bea
\int_{\lambda_o}^{\lambda_s} \d\lambda\delta \theta(\bm n_\text{obs},x(\lambda))
&=&
2\frac{\delta d_A}{\chi}\bigg|_{ x_{{s}}}
-2\delta x^i\nabla_i\frac{\delta d_A}{\chi}
+\int_{\lambda_o}^{\lambda_s} \d\lambda\delta x^i\nabla_i\delta \theta(\bm n_\text{obs},\bar x(\lambda))\\
&=&
2\frac{\delta d_A}{\chi}\bigg|_{ x_{{s}}}
+2(\nabla_i\delta x^i)\frac{\delta d_A}{\chi}
+\int_{0}^{\chi_s} \d\chi(\nabla_i\delta x^i)\delta \theta(\bm n_\text{obs},\bar x(\lambda))\\
&=&2\frac{\delta d_A}{\chi}\bigg|_{ x_{{s}}}
+2(\nabla_i\delta x^i)\frac{\delta d_A}{\chi}
+(\nabla_i\delta x^i)\int_0^{\chi_s}\d\chi \delta\theta
-\int\d\chi \left[\int_0^{\chi}\d\chi' \delta\theta(\chi')\right]\frac{\d}{\d\chi}\nabla_i\delta x^i\\
&=&
2\frac{\delta d_A}{\chi}\bigg|_{ x_{{s}}}
+2\int\d\chi \frac{\delta d_A}{\chi}\frac{\d}{\d\chi}\nabla_i\delta x^i\label{dsjhbc}
\eea
Now, evaluating $\nabla_i\delta x^i$:
\be
\frac{\d}{\d\chi}\nabla_i\delta x^i = \frac{\d}{\d\chi}(N^{ij}\nabla_i\delta x_j)=-\delta\theta
+ \mbox{ subdominant terms}
\label{djkshb}
\ee
which implies
\be
\nabla_i\delta x^i = - \int_0^\chi\d\chi\delta\theta = 2\frac{\delta d_A}{\chi} + \mbox{ subdominant terms.}
\ee
Ignoring subdominant terms we then have 
\be
\int_{\lambda_o}^{\lambda_s} \d\lambda\delta \theta(\bm n_\text{obs},x(\lambda))
=2\frac{\delta d_A}{\chi}\bigg|_{ x_{{s}}} + 2\left(\frac{\delta d_A}{\chi}\right)^2\label{1}
\ee
Note that if we had not employed the shift, the formula would read, from \eqref{dhjsb}, 
\be
\int_{\lambda_o}^{\lambda_s} \d\lambda\delta \theta(\bm n_\text{obs},x(\lambda))
=2\frac{\delta d_A}{\chi}\bigg|_{ \tilde x_{{\bar s}}} - 2\left(\frac{\delta d_A}{\chi}\right)^2\label{2}
\ee
Both formulas are correct, but they mean slightly different things.

To identify which we want to use in a given situation, let us expand the left hand side of~\eqref{sdajhb}:
\be
2\ln d_A = 2\ln\chi_s +2\frac{\delta d_A}{\chi_s}  + \frac{\delta^2d_A}{\chi_s}-\left(\frac{\delta d_A}{\chi_s}\right)^2\ee
The meaning of the second-order term changes depending on whether we choose to evaluate the linear term at the apparent position or the true position of the source. 
Now, if we choose to evaluate the linear term at the physical position of the source we have to use \eqref{1}:
\be
+2\frac{\delta d_A}{\chi_s}\bigg|_{x_s}  + \frac{\delta^2d_A}{\chi_s}-\frac{\delta d_A^2}{\chi_s^2} = 2\frac{\delta d_A}{\chi}\bigg|_{ x_{{s}}} + 2\left(\frac{\delta d_A}{\chi_s}\right)^2
\ee
the linear terms cancel, and so we derive:
\be
\frac{\delta^2d_A}{\chi_s}=3 \left(\frac{\delta d_A}{\chi_s}\right)^2~~~~\text{if $\delta d_A$ is evaluated at the physical position $S$,}\label{3}
\ee
in agreement with~\eqref{e:A9}

On the other hand, if we choose to evaluate the terms in $d_A$ at the background position of the source we would use \eqref{2} to write:
\be
+2\frac{\delta d_A}{\chi_s}\bigg|_{\tilde x_{\bar s}}  + \frac{\delta^2d_A}{\chi_s}-\frac{\delta d_A^2}{\chi_s^2} = 2\frac{\delta d_A}{\chi}\bigg|_{\tilde x_{\bar s}} - 2\left(\frac{\delta d_A}{\chi_s}\right)^2
\ee
from which we have:
\be
\frac{\delta^2d_A}{\chi_s}=- \left(\frac{\delta d_A}{\chi_s}\right)^2~~~~\text{if $\delta d_A$ is evaluated at the apparent position of the source}
\ee
However, the apparent position of the source is a perturbed quantity and a first order 
perturbation evaluated there contains a second order contribution such that its enable average is not guaranteed to vanish.

\noindent {\bf Along geodesic 2:} We now repeat the calculation using geodesic 2 as our reference geodesic:
\bea
\int_{\lambda_o}^{\lambda_s} \d\lambda\delta \theta(\bm n_\text{obs},x(\lambda))&=&
\int_{\lambda_o}^{\lambda_s} \d\lambda\delta \theta(\tilde{\bm n},\bar x_2(\lambda))+\int_{\lambda_o}^{\lambda_s} \d\lambda\widetilde{\delta x}^i\nabla_i\delta \theta(\bm n_\text{obs},\bar x(\lambda))\\
&=&
2\frac{\delta d_A}{\chi}\bigg|_{\tilde{\bm n}}
+\int_{\lambda_o}^{\lambda_s} \d\lambda\widetilde{\delta x}^i\nabla_i\delta \theta(\bm n_\text{obs},\bar x(\lambda))\label{iorjg}
\eea
In the first term we need the distinction between $\tilde n$ and $n_\text{obs}$, in the second, which is already second order this is not necessary. The first term is a function evaluated in terms of an unobservable direction $\tilde{\bm n}$ to the actual source location. To evaluate the second term we use the following relations which are follow from  Fig~\ref{fig:Chris}
\bea
\bar x_1+\delta x&=&\bar x_2+\widetilde{\delta x}\\
\widetilde{\delta x}&=&\delta x-\chi\delta n=\delta x-\frac{\chi}{\chi_s}\delta x_s
\eea 
where the last step follows by similar triangles. Substituting this into \eqref{iorjg} we find,
\be
\int_{\lambda_o}^{\lambda_s} \d\lambda\delta \theta(\bm n_\text{obs},x(\lambda))=2\frac{\delta d_A}{\chi}\bigg|_{\tilde{\bm n}}+ 2\left(\frac{\delta d_A}{\chi}\right)^2\label{dsjhvb}
\ee
in agreement with \eqref{1}. Switching from $\tilde{\bm n}$ to $\bm n_\text{obs}$ reproduces \eqref{2}.

This derivation makes it clear  that the formula in Appendix A applies \emph{only when the linear term is evaluated at the physical position of the source}. When we evaluate ensemble averages this is the formula we need to use. The apparent position of the source moves in response to lensing events whereas the physical position does not. Thus, the term which `appears' as linear when we use the observed apparent position of the source, $\bar S$, in fact has second-order corrections which can move $\bar S$ around. 

\section{Derivation of \eqref{jdshbcsjhbca} from the lensing map}\label{derivation3}

Let us start with the lens map~\cite{CamilleLens},
\be
\mathcal{D}= \la \left(\begin{array}{cc} 1-\kappa-\gamma_1 & -\gamma_2 \\  -\gamma_2 & 1-\kappa+\gamma_1 \end{array}\right) \,,
\ee
where $\kappa$ denotes the convergence, $\gamma = \gamma_1 +i\gamma_2$ is the complex shear and $\la$ is the affine parameter of the photon geodesic.
We perturb all quantities to 2nd order
\bea
\la &=& \chi + \la^{(1)} + \frac{1}{2}\la^{(2)} \\
\kappa(S) &=&  \kappa^{(1)}(\bar S) + \frac{1}{2}\kappa^{(2)}(\bar S)  \\
\gamma_{1,2}(S) &=&\gamma^{(1)}_{1,2}(\bar S) + \cdots  
\eea 
Here we indicate  how perturbed quantities are understood. Namely the true $\kappa$ at the true source position $S$ is approximated by a 1st and second order term both evaluated at the image  position $\bar S$, see Fig.~\ref{fig:Chris}.
 
As we have argued before, perturbations of the affine parameter $\la$ are subdominant.
Up to second order we then have
\be
d_A^2(S) =\det (\mathcal{D}) = \chi^2\left[ 1-2 \kappa^{(1)}(\bar S)  - \kappa^{(2)}(\bar S) +  (\kappa^{(1)} )^2 (\bar S)
-(\gamma^{(1)}_1)^2(\bar S) -(\gamma^{(1)}_2)^2(\bar S)\right] = \chi^2\left[ 1-2 \kappa^{(1)}(\bar S) 
+ t(\bar S)\right] \,,
\ee
where $t$ is a total divergence. This  can be verified, e.g. using the second order lensing expressions which have been calculated in Ref.~\cite{CamilleLens}. Let us take the square root:
\be\label{Dx}
d_A(S) =  \chi\left[ 1- \kappa^{(1)}(\bar S) -\frac{1}{2} \left(\kappa^{(1)}(\bar S)\right)^2 + 
\frac{1}{2} t(\bar S)\right] \,.
\ee
We want to determine the 2nd order contribution to this quantity. For this we subtract the
0th and 1st order. But to subtract  the 1st order we have to evaluate both sides at the same position. To 2nd order this does make a difference.
\be 
d_A(S) = \chi\left[ 1 + \frac{\de d_A(S)}{\chi}+ \frac{1}{2} \frac{\de^2 d_A(S)}{\chi} \right]
\ee
Expanding also 
\be
 \kappa^{(1)}(\bar S)  =  \kappa^{(1)}(S) -\de x^a\nabla_a\kappa^{(1)}(\bar S) 
 =  \kappa^{(1)}(S)+ (\nabla_a\de x^a)\kappa^{(1)}(\bar S) 
=  \kappa^{(1)}(S) - 2[\kappa^{(1)}(\bar S)]^2 + \mbox{total divergence}
\ee
Here we have used $ (\nabla_a\de x^a)= -2\kappa^{(1)}$ to first order.

Now we evaluate both sides as the same position and can safely subtract the 0th and 1st order terms.  With the first order identity
$$ -\kappa^{(1)} (S) = \frac{\de d_A(S)}{\chi} $$
we finally obtain
\be
\frac{\de^2 d_A(S)}{\chi} =3(\kappa^{(1)} )^2(\bar S) +\mbox{ total divergence}  = 3\frac{ (\de d_A)^2}{\chi^2} +\mbox{ total divergence.} 
\ee

\section{Proof that the area of the CMB is larger with structure present.}

The area of a surface of constant redshift is the integral of $d_A^2$ over the observers sky. Let us consider $d_A^2$. In general this obeys:
\be
k^a\nabla_a d_A^2 = \theta d_A^2\,.
\ee
Written relative to the distance in the background we can write this as (to all orders in perturbations)
\be
k^a\nabla_a \left(\frac{d_A}{\chi}\right)^2 = \left(\theta-\frac{2}{\chi} k^a\nabla_a\chi\right) \left(\frac{d_A}{\chi}\right)^2\,.
\ee
Now let us expand the derivative into a background and total perturbation part:
\be
k^a\nabla_a=-\frac{\d}{\d\chi}+b\Delta k^a\nabla_a\,.
\ee
We have made no approximation here, we just write $k^a=\bar k^a +\De k^a$. The $b$ is just a constant to keep track of the post-Born terms, $b=1$ includes them, $b=0$ just integrates on the background geodesic.
Setting
\be
\left(\frac{d_A}{\chi}\right)^2=1+\frac{\Delta d_A^2}{\chi^2}~~~ \text{and} ~~~\Delta\theta=\theta+\frac{2}{\chi}\,.
\ee
 we can write
\be
\frac{\d}{\d\chi}\frac{\Delta d_A^2}{\chi^2}
=-\Delta\theta\left(1+\frac{\Delta d_A^2}{\chi^2}\right)+b\Delta k^a\nabla_a\frac{\Delta d_A^2}{\chi^2}\,.
\ee
So far, these are just definitions, not yet implying approximations (though all $\Delta$ quantities vanish on the background).
Now we can start applying approximations used above. Up to subdominant terms $\nabla_a\Delta k^a = \De\theta$. Therefore up to a total divergence and subdominant terms,
\be
\Delta k^a\nabla_a\frac{\Delta d_A^2}{\chi^2}=-\Delta\theta \frac{\Delta d_A^2}{\chi^2}\,
\ee
neglecting lower derivative terms from the perturbed Christoffel symbols. With this 
\be
\frac{\d}{\d\chi}\frac{\Delta d_A^2}{\chi^2}+(1+b)\Delta\theta\frac{\Delta d_A^2}{\chi^2}
= -\Delta\theta\,.
\ee
Integrating,
\be
\frac{\Delta d_A^2}{\chi^2}=-e^{-(1+b)\int^{\chi}\d\chi'\Delta\theta}\int\d\chi'
\Delta\theta e^{(1+b)\int^{\chi'}\d\chi''\Delta\theta}\,.
\ee
Integrating this result by parts, we find
\be
\frac{\Delta d_A^2}{\chi^2}=-\int_0^\chi\d\chi'\Delta\theta+(1+b)e^{-(1+b)\int^{\chi}\d\chi'\Delta\theta}\int\d\chi'\left(\int^{\chi'}\d\chi''\Delta\theta\right)
\Delta\theta e^{(1+b)\int^{\chi'}\d\chi''\Delta\theta}\,.
\ee
The exponential terms only affect the solution at 3rd order, so we can ignore them. Using $\int\d\chi'\left(\int^{\chi'}\d\chi''\Delta\theta\right)
\Delta\theta=\frac{1}{2}\left(\int^{\chi}_0\d\chi'\Delta\theta\right)^2$, we arrive at
\be
\frac{\Delta d_A^2}{\chi^2}=-\int_0^\chi\d\chi'\Delta\theta+\frac{1+b}{2}\left(\int^{\chi}_0\d\chi'\Delta\theta\right)^2\,.
\ee
The first term always forms a total divergence order-by-order, while the second term \emph{always gives a positive contribution to the area}.

\section{Derivation of \eqref{ljshdbjasdhb} }\label{A:5}
We start with \eqref{jdshbcsjhbca}: 
\bea
\Delta_\text{int}(\chi_s,\bm n)&= &\frac{3}{2}\int_0^{\chi_s}{\d\chi_1}\frac{(\chi_s-\chi_1)}{\chi_s\chi_1}\int_0^{\chi_s}{\d\chi_2}\, \frac{(\chi_s-\chi_2)}{\chi_s\chi_2}\De_2\Phi(\chi_1)\De_2\Phi(\chi_2)
\eea
Expanding $\Phi$ in spherical harmonics with \eqref{kjsbdnkjsd}, taking an ensemble average and performing one $k$-integral gives
\bea\label{sakjncsdacjkn}
&&\langle\Delta_\text{int}\rangle= 6\pi\int_0^{\chi_s}\!\d\chi_1\frac{(\chi_s-\chi_1)}{\chi_s\chi_1}\int_0^{\chi_s}\!\d\chi_2\frac{(\chi_s-\chi_2)}{\chi_s\chi_2}g(\chi_1)g(\chi_2)
\sum_{\ell_1 m_1}\sum_{\ell_2 m_2}i^{\ell_1-\ell_2}\!\int\! \frac{\d k}{k}\mathcal{P}_0T(k)^2 j_{\ell_1}(k\chi_1)j_{\ell_2}(k\chi_2)\\&&\nonumber
~~~~~~~~~~~~~~~~~ \times\int\d\Omega_K Y_{\ell_1 m_1}(\hat{\bm k})Y^*_{\ell_2 m_2}(\hat{\bm k})
\ell_1(\ell_1+1)\ell_2(\ell_2+1)
Y_{\ell_1 m_1}(\bm n)
Y^*_{\ell_2 m_2}(\bm n)
\,.
\eea
Now we can perform the angular integral in $k$-space and use
$
\int\d\Omega_k Y_{\ell_1 m_1}(\hat{\bm k}) Y^*_{\ell_2 m_2}(\hat{\bm k})=\delta_{\ell_1\ell_2}\delta_{m_1m_2}$, and from
$
\sum_{m=-\ell}^{m=\ell} Y_{\ell m}(\bm n)Y^*_{\ell m}(\bm n)=({2\ell+1})/{4\pi}
$
we derive 
\bea
\langle\Delta_\text{int}\rangle= \frac{3}{2}\int_0^{\chi_s}{\d\chi_1}\frac{(\chi_s-\chi_1)}{\chi_s\chi_1}\int_0^{\chi_s}{\d\chi_2}\frac{(\chi_s-\chi_2)}{\chi_s\chi_2}g(\chi_1)g(\chi_2)
\nonumber
\sum_{\ell=0}^\infty(2\ell+1)[\ell(\ell+1)]^2\int \frac{\d k}{k}\mathcal{P}_0T(k)^2 j_{\ell}(k\chi_1)j_{\ell}(k\chi_2)\,.
\eea
This can be accurately simplified using the Limber approximation:
\be
\int \frac{\d k}{k}
\mathcal{P}_0T(k)^2 j_{\ell}(k\chi_1)j_{\ell}(k\chi_2)\simeq
\frac{4\pi\chi_1}{(2\ell+1)^3}\mathcal{P}_0T^2\left(k=\frac{\ell+1/2}{\chi_1}\right)\delta(\chi_1-\chi_2)\,,
\ee
so
\bea
\langle\Delta_\text{int}\rangle&=& 6\pi\sum_{\ell=0}^\infty\left[\frac{\ell(\ell+1)}{2\ell+1}\right]^2\int_0^{\chi_s}\d\chi\frac{(\chi_s-\chi)^2}{\chi\chi_s^2}g(\chi)^2
\mathcal{P}_0 T^2\left(k=\frac{\ell+1/2}{\chi}\right)\,.
\eea
This is \eqref{ljshdbjasdhb}.\\[5mm]

\section{Local terms at low redshift}\label{lsadvbasdb}

For our purposes the most important local terms are the ones with the highest even numbers of spatial derivatives  (odd numbers of spatial derivatives do not contribute to an ensemble average):
\bea
\Delta_\text{loc}&=&\left(\frac{\HH'}{2\HH^2}-\frac{1}{\chi_s\HH_s}-\frac{1}{2}\right)\!\!\D_{\pp}v_s\D_{\pp}v_s+\chi_s\D_{\pp}\Phi_s\D_{\pp}v_s\nonumber\\&&
\hspace{-1.5cm}+
\left(1-\frac{1}{\chi_s\HH_s}\right)\!\!
\big[\chi_s\D_{\pp} v_s \D_{\pp}v'_s +\nabla_{\o i}v_s\nabla_{\o}^iv_s
+\chi_s\Phi_s\D_{\pp}^2v_s\big]\!.~~~~\label{dloc}
\eea
{Here {$v_s$} is the velocity potential at the source, and $\D_{\pp}, \nabla_{\o i}$ are the radial and transverse (screen-space) parts of the spatial derivative.
This contains} a mixture of radial and transverse velocity terms, as well as the redshift-space distortion effect.

For $\langle\Delta_\text{loc}\rangle$, all terms in \eqref{dloc} are proportional to $(\nabla\p\Phi_0)^2$, {so $\langle\Delta_\text{loc}\rangle\propto\langle (\nabla\p\Phi_0)^2\rangle$, where}
\bea
\langle (\nabla\p\Phi_0)^2\rangle 
={\frac{1}{2}} \langle\nabla_{\o i}\Phi_0\nabla_\o^i\Phi_0\rangle
=\frac{1}{3}{\int\d k\, k \mathcal{P}_0(k) T^2(k)}.
\eea
Summing up the terms in \eqref{dloc} results in a small negative contribution, and is the dominant contribution at low-$z$ ($z\lesssim0.5$), {as shown in Fig.~\ref{sdjhbcsdb}}.

\end{widetext}

\end{document}